\documentclass[fleqn,twoside]{article}
\usepackage{espcrc2}

\usepackage{graphicx}
\usepackage[figuresright]{rotating}

\newcommand{\AmS}{{\protect\the\textfont2
  A\kern-.1667em\lower.5ex\hbox{M}\kern-.125emS}}

\newcommand{\newc}{\newcommand}
\newc{\lra}{\leftrightarrow}
\newc{\beq}{\begin{equation}}
\newc{\eeq}{\end{equation}}
\newc{\barr}{\begin{eqnarray}}
\newc{\earr}{\end{eqnarray}}
\title{Neutrinoless Double Beta Decay in Theories Beyond the Standard Model}
\author{J.D. Vergados\address{University of Ioannina, Gr 4510 10,
Ioannina, Greece}}
\begin{document}
\begin{abstract}
Neutrinoless double beta decay pops up almost in any extension of
the standard model. It is perhaps the only process, which can
unambiguously determine whether the massive neutrinos are Majorana
or Dirac type particles. In addition from the lifetime of this
decay, combined with sufficient knowledge of the relevant nuclear
matrix elements, one can set a constraint involving the neutrino
masses. Furthemore, if one incorporates the recent results of the
neutrino oscillation experiments, one can determine or set a
stringent limit on the neutrino mass scale. In addition one may
obtain usefull information regarding the presence of right handed
currents and the right handed neutrino mass scale. One can also
constrain the parameters of supersymmetry and, in particular, set
limits in of R-parity violating couplings as well as get
information about extra dimensions.
 \vspace{1pc}
\end{abstract}
 \maketitle
\section{THE INTERMEDIATE NEUTRINO MECHANISM}
Neutrinoless double beta decay occurs whenever ordinary beta decay
is forbidden due to energy conservation or retarded due to angular
momentum mismatch, but the nucleus, which is two units of charge
away, is accessible via a order week
interaction\cite{Ver02}-\cite{FKSS97}.  It is a process known for
almost 70 years, which has been searched for, but not seen yet. It
is still of great theoretical and experimental interest since
perhaps it is the only process, which can unambiguously determine
whether massive neutrinos can be Majorana or Dirac type particles.
It can occur only if the mass eigenstates are
 Majorana particles (the particle coincides with its antiparticle).

 In the weak basis $\nu^0$, $\nu^{0c}$ the neutrino mass matrix
 takes the form:
 $\left ( \begin{array}{c}\bar{\nu}^0_L \\
 \bar{\nu}^{0c}_L \\
\end{array} \right )\left ( \begin{array}{cc}
m_{\nu}&m_{D}\\
m^T_{D}&m^{N}\\
 \end{array} \right )
\left ( \begin{array}{c}\nu^{0c}_ R\\
\nu^0_R\end{array} \right )$\\
where $m_{\nu}$ is a $3 \times 3$ left-handed (isotriplet)
Majorana mass matrix, $m_D$ is a Dirac $3 \times N$ mass matrix
and $m_N$ is the right-handed (isosinglet) $N \times N$ Majorana
mass matrix. In models motivated by GUT's $N=3$, $m_{\nu}=0$,
$m_D$ is of the order of the up quark mass matrix and $m_N$ is
very heavy ($\geq 10^{10}$ GeV) so that light neutrinos with mass
of order $m_D(m_n)^{-1}m^T_D$ can occur. On the other hand in
R-parity violating supersymmetry only $m_{\nu}$ is non zero.
Anyway after diagonalizing the above matrix the weak eigenstates
are related to the mass eigenstates via the equation:
$\left ( \begin{array}{c}\nu^0_L \\
 \nu^{0c}_L \\
\end{array} \right )= \left ( \begin{array}{cc}
U_{11}&U_{12}\\
U^{21}&U^{22}\\
 \end{array} \right )
\left ( \begin{array}{c}\nu^c_ R\\
\nu_R\end{array} \right )$\\
In the above notation $U^{(11)}=U_{MNS}$ is the usual Maki,
Nakagawa, Sakata charged leptonic matrix, which for convenience
sometimes will merely be indicated by $U$.

 \subsection{Left-handed currents only}
\begin{figure}
\hspace*{-0.0 cm}
\includegraphics[height=.2\textheight]{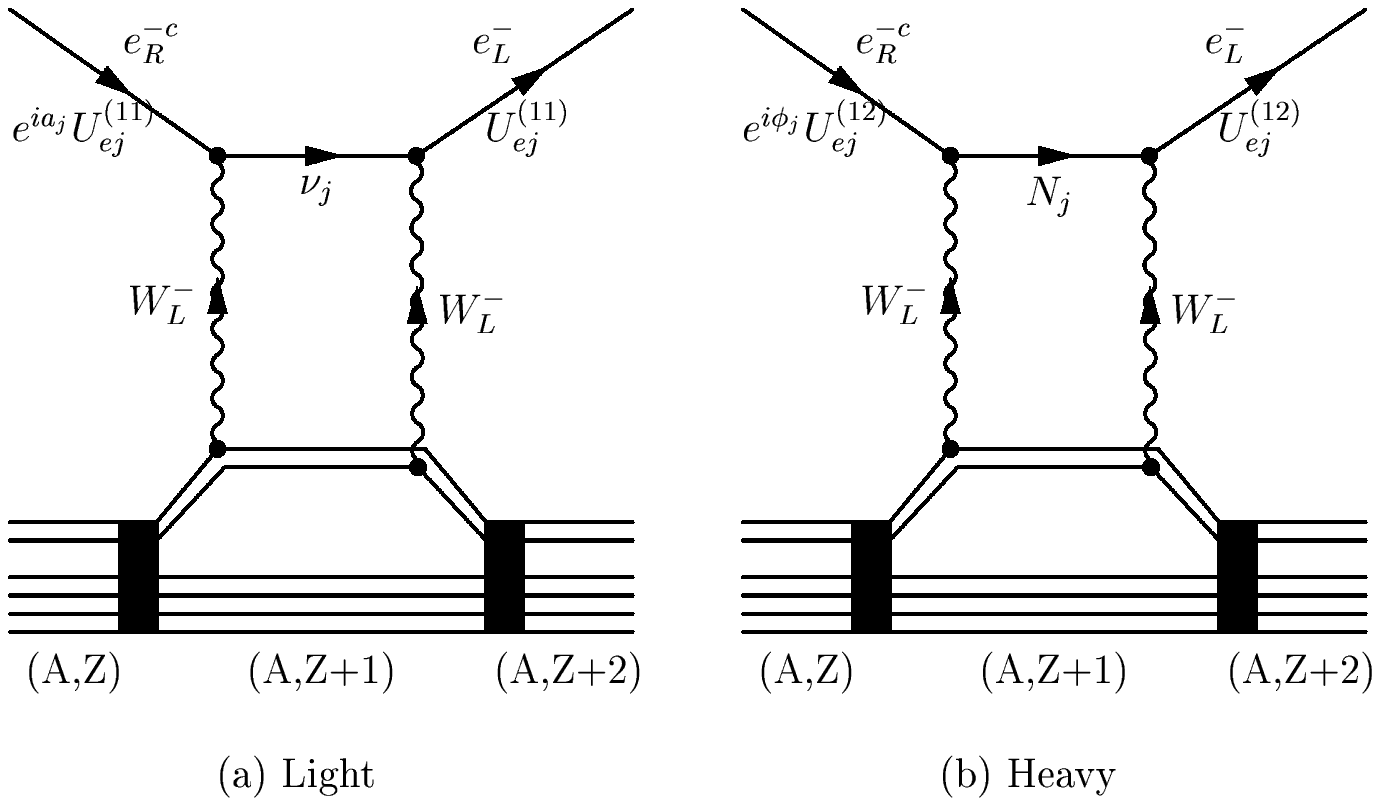}
\caption{neutrinoless double beta decay in left handed theories
for light (a) and heavy (b) intermediate neutrinos. \label{lmass}}
 \end{figure}

  In the presence of only left handed currents one has a
contribution
 to neutrinoless double beta decay arising  from the diagram
  shown in fig. \ref{lmass}.
One encounters two lepton violating parameters $ \eta_{\nu}$ and
$\eta^L_{_N}$ given by \cite{Ver02}:
 \beq
 \eta_{\nu}=<m_\nu>/m_e
 \label{lmasse1a}
 \eeq
 \beq
  <m_\nu > ~ = ~ \sum^{3}_j~ (U^{(11)}_{ej})^2 ~ \xi_j ~
m_j,
 \label{lmasse1}
    \eeq
     \beq
 \eta^L_{_N} ~ = ~ \sum^{3}_j~ (U^{(12)}_{ej})^2 ~
    ~\Xi_j ~ \frac{m_p}{M_j}
    \label{lmasse}
    \eeq
with $m_j$ ($M_j$) the light (heavy) neutrino masses and
$\xi_j=e^{i \alpha_j}$ ($\Xi_j=e^{i \Phi_j}$) the Majorana phases
of the corresponding mass eigenstates. The separation of these
particle physics parameters from the nuclear physics holds, if the
neutrino is very light ($m_j<<m_e$) or  much heavier than the
proton ($M_j>>m_p$).
\subsection{The leptonic right-left interference contribution}
In the presence of right-handed currents one can have a
contribution via light intermediate neutrinos, in which the
corresponding amplitude does not explicitly depend on the neutrino
mass \cite{Ver02}(see Fig. \ref{lnomass}). We encounter two lepton
violating parameters:
  \begin{figure}
\hspace*{-0.0 cm}
\includegraphics[height=.4\textheight]{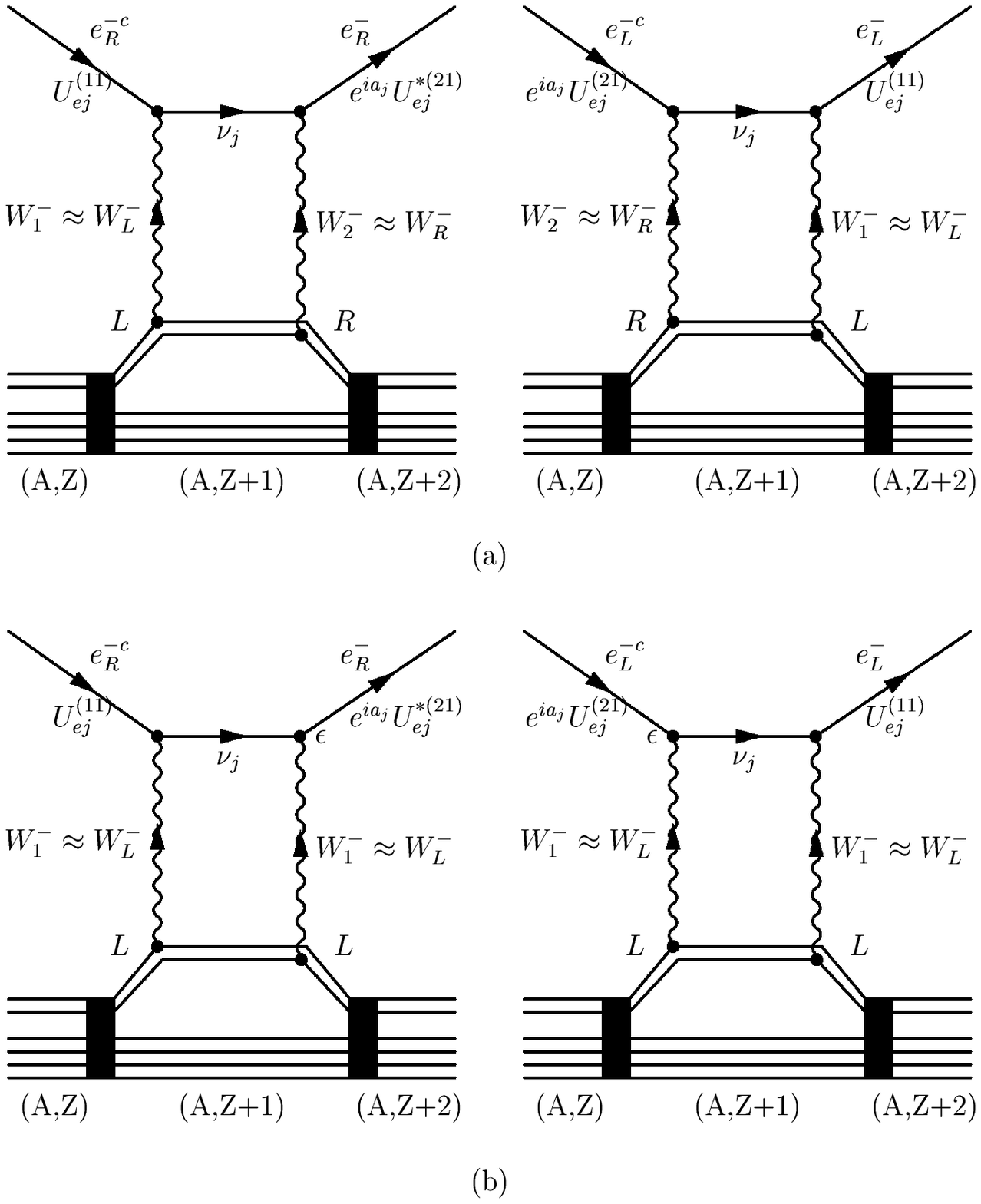}
\caption{Diagrams leading to neutrinoless double beta decay with
amplitude not explicitly dependent on the neutrino mass.
 \label{lnomass}}
 \end{figure}
 \beq
\eta ~ = ~ \epsilon \eta_{RL} , \lambda ~=~ \kappa
\eta_{RL},\eta_{RL} ~=~
     \sum^{3}_j~ (U^{(21)}_{ej}U^{(11)}_{ej}) ~\xi_j
     \label{rle}
     \eeq
     where $\kappa=m^2_L/m^2_R~~,~~\epsilon=tan \zeta$.
 $m_L,m_R$ are
the gauge boson masses and $\zeta$ the gauge boson mixing angle.
\subsection{Right-handed neutrino mass term}
\begin{figure}
\hspace*{-0.0 cm}
\includegraphics[height=0.2\textheight]{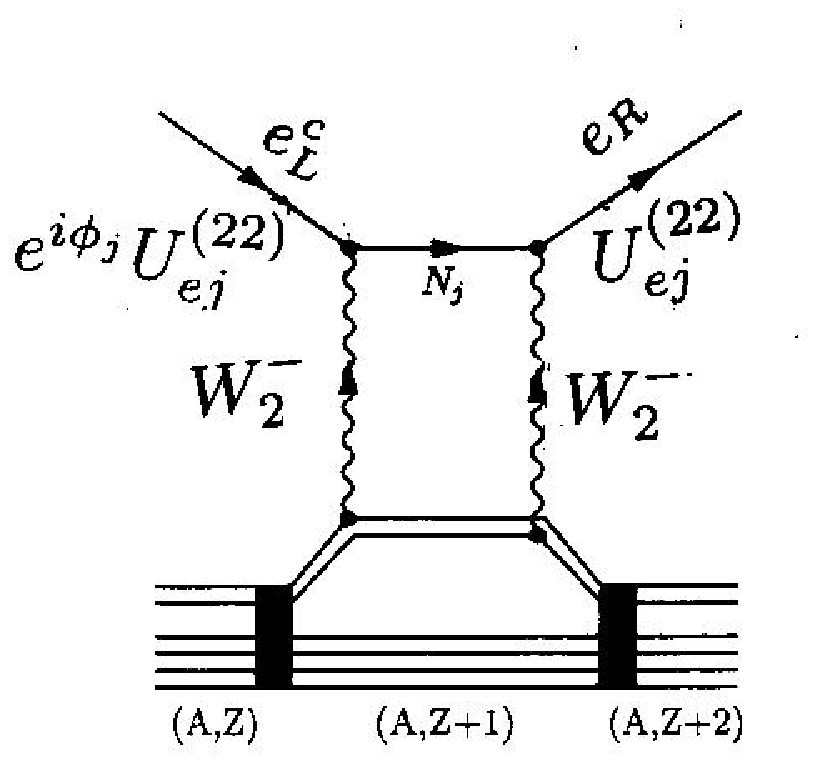}
 \includegraphics[height=0.15\textheight]{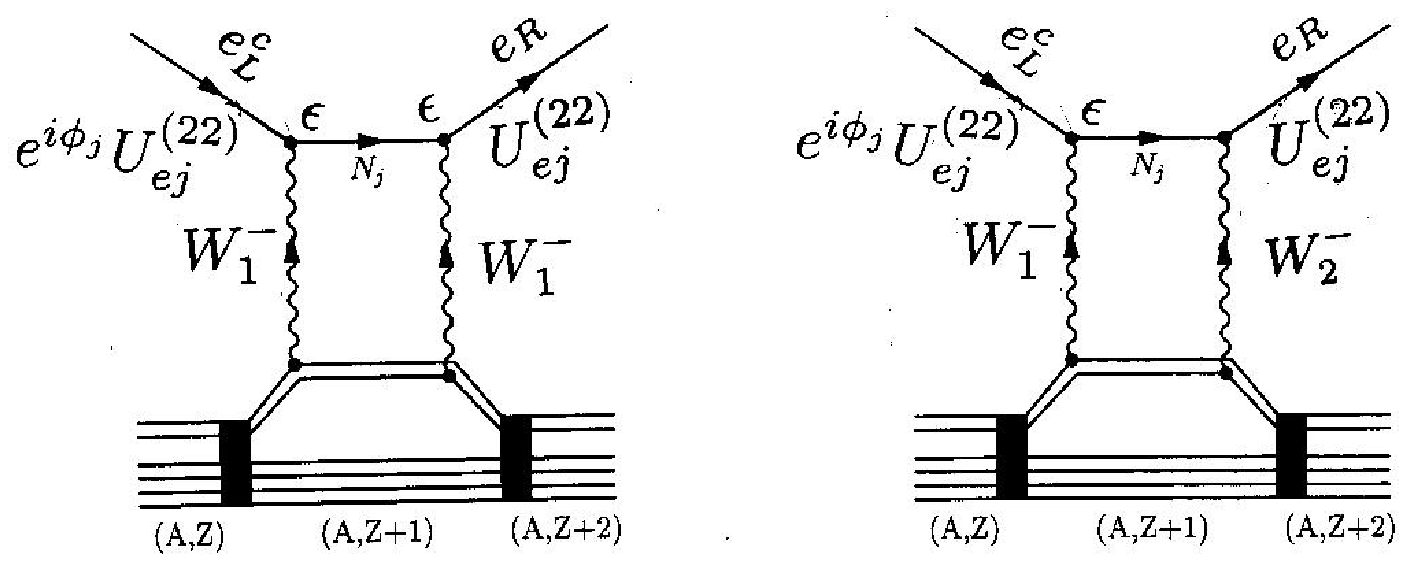}
 \caption{The heavy neutrino contribution due to the right handed
 current\label{hmass}}
\end{figure}

 In the presence of right-handed currents one can have additional
  mass dependent  terms (see \ref{hmass}).
  The corresponding lepton violating parameter is given by:
  \beq
   \eta^R_{_N} ~ =~(\kappa ^2 + \epsilon ^2+c_0\epsilon \kappa) \sum^{3}_j~ (U^{22}_{ej})^2 ~
    ~\Xi_j ~ \frac{m_p}{M_j}
    \label{rmasse}
    \eeq

\subsection{Analysis of the data in terms of light neutrinos}
In the presence of right handed currents the neutrinoless double
beta decay lifetime is given \cite{Ver02}, \cite{PSV96} by:
  \begin{eqnarray}
 [T_{1/2}^{0\nu}]^{-1} &=& G^{0\nu}_{01} |M^{0\nu}_{GT}|^2 [
|X_{L}|^2  + {\tilde C}_{0\lambda} |\lambda| X_{L} cos \psi_1 +
 \nonumber\\
& & {\tilde C}_{0\eta} |\eta| X_{L} cos \psi_2 + {\tilde
C}_{\lambda\lambda}|\lambda|^2 +{\tilde C}_{\eta\eta}+
 \nonumber\\
  & &|X_R|^2+
|\eta|^2   + {\tilde C}_{\lambda\eta}|\lambda||\eta| cos (\psi_1
-\psi_2)]
\nonumber\\
& &+|X_R|^2+ Re ({\tilde C}_{0\lambda} \lambda X_{R} + {\tilde
C}_{0\eta} \eta X_{R})], \label{eq:70}
 \end{eqnarray}
  where
 $$ X_{L}^{} = \frac{<m_\nu>} {m_{e}}
(\chi_{F}^{}-1) +\eta^L_N \chi^{}_{H}+..., X_{R} =\eta^R_N
\chi^{}_{H}+...$$
$$ \chi^{}_{F} = (\frac {g^{}_{V}}{g^{}_{A}})^2 \frac
{M_{F}(0\nu)(0\nu)}{M_{GT}}$$ $$\chi^{}_{H} = ((\frac
{g^{}_{V}}{g^{}_{A}})^2 M^{}_{FH} - M^{}_{GTH})/ M^{}_{GT}$$
 where the subscript $H$ indicates heavy particle (neutrino),
 $\psi_1$ is the relative phase between $X_{L}$ and $\lambda$ and
 $\psi_2$ is the relative phase between $X_{L}$ and $\eta$, while
  $\tilde{C}_i$ depend on the energy and nuclear physics
 The ellipses \{...\} indicate contributions
arising from other particles such as intermediate SUSY particles,
unusual particles predicted by superstring models, unusual
messengers in extra dimensions, exotic Higgs scalars etc.

\begin{table}[t]
 \caption{Summary of the present experimental
results.}
\begin{center}
\bigskip
\label{table.8}
\begin{tabular}{|l|r|l}
 \hline
 & &   \\
 Isotope            & T$_{1/2}^{0\nu}$ (y) &Refs\\
 \hline
$^{48}$Ca          & $>9.5\times 10^{21} (76\%)$                                   & \cite{YOU91} \\
$^{76}$Ge          & $>1.9\times 10^{25}(90\%)$                                          & \cite{KLA01} \\
                   & $>1.6\times 10^{25}(90\%)$                                         & \cite{AAL99}  \\
$^{82}$Se          & $>2.7\times 10^{22}(68\%)$                                  & \cite{ELL92}  \\
$^{100}$Mo         & $>5.5\times 10^{22}(90\%)$                                               & \cite{EJI96}  \\
$^{116}$Cd         & $>7\times 10^{22}(90\%)$                                                & \cite{DAN00}  \\
$^{128,130}$Te     & $R= (3.52\pm 0.11)\times 10^{-4}$   &\cite{BER93} \\
                   &(geochemical)                                                                &                 \\
$^{128}$Te         & $>7.7\times 10^{24}(90\%)$                                       &\cite{BER93} \\
$^{130}$Te         & $>1.4\times 10^{23}(90\%)$                                             & \cite{ALE00}\\
$^{136}$Xe         & $>4.4\times 10^{23}(90\%)$                                          & \cite{LUE98}\\
$^{150}$Nd         & $>1.2\times 10^{21}(90\%)$
$<3(90\%)$                             & \cite{DES97} \\
\end{tabular}
\end{center}
\end{table}

\begin{figure}
 \hspace*{-0.0cm}
\includegraphics[height=.5\textheight]{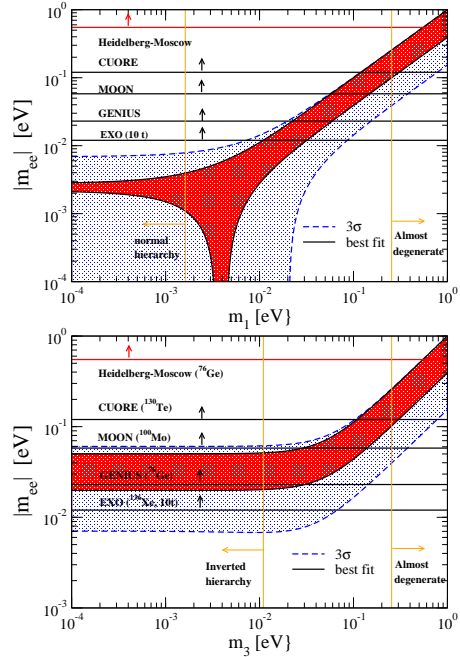}
%
 \caption{ The lightest neutrino mass obtained from
 $0\nu \beta \beta$ decay and  neutrino oscillation
 data for normal (top) and inverted (bottom) hierarchy.
  \label{mlightest}}
\end{figure}

   As we have already mentioned the experiment (see table \ref{table.8}) imposes
 only one constraint among the lepton
  violating parameters.
 Normally one assumes that only one lepton
  violating parameter dominates and proceeds extracting a limit on that
  one (see e.g. Table \ref{table.9} below). In the most favored scenario one assumes that
  the light neutrino mass mechanism dominates. One can then employ the the
  neutrino oscillation data \cite{BILENKY} to get the constraints of Fig. \ref{mlightest}.
  Clearly $0\nu \beta \beta$ decay can differentiate between the hierarchies.
  In fact the degenerate scenario supports the recent claims
  that $0\nu \beta \beta$ decay  may even have been seen \cite{KLAP04}.
  In the inverted hierarchy, $0\nu \beta \beta$  can even set a limit on the
  lightest neutrino mass $m_1$.

\begin{figure}
\rotatebox{90}{\hspace{1.0cm} $<m_{\nu}>\rightarrow$ eV}
\includegraphics[height=.18\textheight]{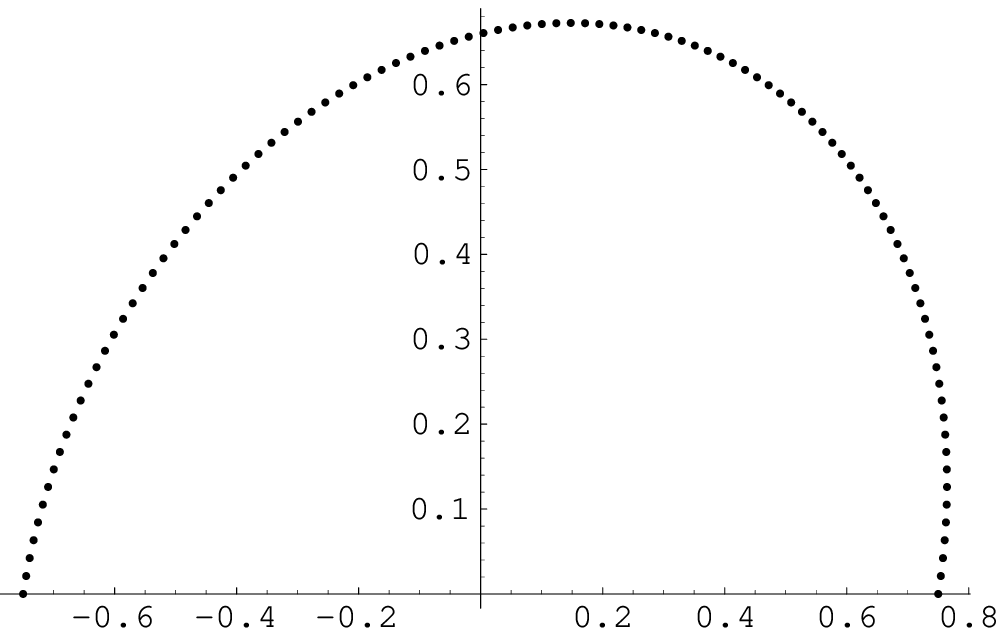}
\hspace{0.0cm} {\tiny $\lambda \times 10^{-6}$}
 \caption{The excluded region in the space of
  the lepton violating parameters $<m_{\nu}>$
and $\eta$ entering neutrinoless double beta
decay.
 \label{masseta}}
 \end{figure}

\begin{figure}
 \hspace*{-0.0 cm}
 \rotatebox{90}{\hspace{1.0cm} $\lambda \rightarrow 10^{-6}$}
 \includegraphics[height=.17\textheight]{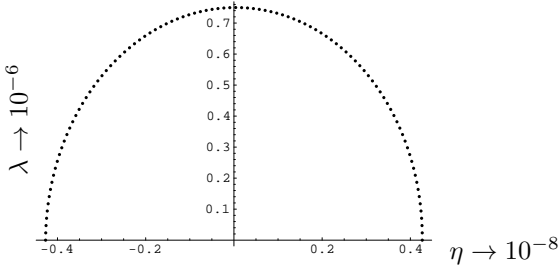}
{\hspace{0.0cm} $\eta \rightarrow 10^{-8}$}
 \caption{The constraint on the $\eta$ $\lambda$ parameters extracted
 from $0\nu \beta \beta$ decay.\label{etalam}}
 \end{figure}

   A more reasonable procedure would
  be to consider multi-dimensional exclusion plots as, e.g., that
  involving $<m_\nu>$ and $\eta$, shown in Fig. \ref{masseta},
   obtained with nuclear matrix elements found in the literature \cite{PSV96}.
  From this plot one extracts the standard limits as the
  intersection of the exclusion curves from the axes.
  One may use neutrinoless double beta decay to extract
  information on the parameters of the right handed interaction
  $\kappa$ and $\epsilon$ from the parameters $\lambda$ and $\eta$
   (see Fig. \ref{etalam}). from the definitions one finds:
  \beq
  \frac{\kappa}{\eta}=\frac{\lambda}{\eta}
  \label{koverl}
  \eeq
Furthermore Eq. \ref{rmasse} can be cast in the form
\cite{Ver02},\cite{PREZEAU}:
   \beq
   (\kappa ^2 + \epsilon ^2+c_0\epsilon \kappa)
   <\frac{m_p}{M_N}>= \eta^R_{_N},c_0 \approx 30
    \label{kl2}
    \eeq
    Where  $ <\frac{m_p}{M_N}>=\sum^{3}_j~ (U^{22}_{ej})^2 ~
    ~\Xi_j ~ \frac{m_p}{M_j}$.
%

\begin{figure}
 \hspace*{-0.0cm}
 \rotatebox{90}{\hspace{1.0cm} $\epsilon \rightarrow$}
\includegraphics[height=.15\textheight]{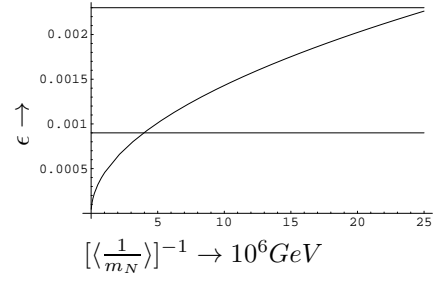}

{\hspace{1.cm}$[\langle\frac{1}{ m_{N}}\rangle]^{-1}\rightarrow
10^{6} GeV $}
 \caption{ The value of $\epsilon$ extracted from $0\nu \beta \beta$
  decay as a function of the average right-handed
 neutrino mass
  \label{eps}}
\end{figure}

\begin{figure}
\hspace*{-0.0 cm}
 \rotatebox{90}{\hspace{1.0cm} $m_{w_R}\rightarrow GeV$}
\includegraphics[height=.15\textheight]{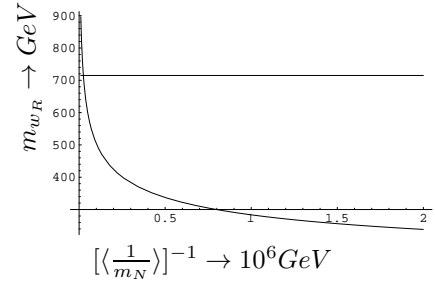}

{\hspace{1.0cm} $[\langle\frac{1}{ m_{N}}\rangle]^{-1}\rightarrow
10^{6} GeV $}
\caption{ The same as in Fig. \ref{eps} for the right handed boson
mass.\label{mwR}}
 \end{figure}

    Using  Eqs (\ref{koverl}) and (\ref{kl2}) one can extract
    limits on $m_{WR}$ and $\epsilon$. Thus, e.g., using:
    $$\lambda=\frac{0.8}{\sqrt{2}}
    \times10^{-6},\eta=\frac{0.4}{\sqrt{2}}
     \times10^{-8},\eta^R_{_N}=0.8\times10^{-8}$$
    we obtain the curves shown in Figs \ref{eps} and \ref{mwR} as a function of the average
    heavy neutrino mass. These should be compared with the values  \cite{PREZEAU}
$$m_{wR}\geq715GeV~,~\epsilon=0.0016\pm0.0007$$
obtained from other experiments.
 \section{$0\nu \beta \beta$ DECAY IN R-PARITY NON -CONSERVING
 SUPERSYMMETRY}
R-parity of supersymmetry is defined as $$R_P=(-1)^{3B+L+2S}$$
 ($B$, $L$ and $S$
are the baryon, lepton numbers and the spin) and is assumed to be
conserved . This allows the R-parity conserving superpotential
     \begin{eqnarray}
        W_{R_p} &= &\lambda^E_{ij} H_1 L_i E^c_j +
                   \lambda^D_{ij} H_1 Q_i D^c_j +
                   \lambda^U_{ij} H_2 Q_i U^c_j +
                   \nonumber\\
                   & & \mu H_1 H_2,
                   \label{super1}
 \end{eqnarray}
 Once R-parity is not conserved one has additional lepton and baryon
 violating terms in the superpotential of the form
  \begin{eqnarray}
  W_{noR_p} & = & \lambda_{ijk}L_{i}L_{j}E^c_{k}
  + \lambda^{\prime}_{ijk}L_{i}Q_{j}D^c_{k}
  +\mu_j L_{j}H_2+
   \nonumber\\
  & & \lambda^{\prime \prime}_{ijk} U^c_{i}D^c_{j}D^c_{k}
  \label{super2}
   \end{eqnarray}
 \begin{eqnarray}
  Q_i & = &
  \left(
  \begin{array}{c}
  u_L \\
  d_L
  \end{array}\right ),\left(
  \begin{array}{c}
  \tilde{u} \\
  \tilde{d}
  \end{array}\right ), U_i^c=u^c_L (\tilde{u}^*), etc
  \nonumber
  \end{eqnarray}
\begin{eqnarray}
  L_i & = &
  \left(
  \begin{array}{c}
  \nu_{eL} \\
  e_L
  \end{array}\right ),\left(
  \begin{array}{c}
  \tilde{\nu}_e \\
  \tilde{e}
  \end{array}\right ), E_i^c=e^c_L(\tilde{e}^*), etc
  \nonumber
   \end{eqnarray}
 One imposes a  discreet symmetry, leading to $\lambda^{\prime}_{ijk}=0$ or
 $\lambda^{\prime \prime}_{ijk}=0$, to
   avoid fast proton decay. In our case we will allow
    for lepton violation imposing $\lambda^{\prime
    \prime}_{ijk}=0$. The bilinear and trilinear R-parity
    violating couplings give rise to massive Majorana neutrinos,
    which in turn contribute to $0\nu \beta \beta$ decay. Here we
    will consider additional contributions.
    The effective Lagrangian describing this process is
    \cite{Ver02}, \cite{FS98}, \cite{FKSS97}
     \begin{eqnarray}
 \nonumber
{\cal L}^{\Delta L_e =2}_{eff} & =& \frac{G_F^2}{2 m_{_p}}~ \bar e
(1 + \gamma_5) e^{\bf c}
 [\eta_{PS}~ J_{PS}J_{PS}
 \nonumber\\
  & & - \frac{1}{4} \eta_T ~
J_T^{\mu\nu} J_{T \mu\nu} + \eta_N ~ J^\mu_{VA} J_{VA \mu}].
 \nonumber
\end{eqnarray}
The color singlet hadronic currents are
$$J_{PS} =   {\bar
u}^{\alpha} (1+\gamma_5) d_{\alpha}, J_T^{\mu \nu} = {\bar
u}^{\alpha} \sigma^{\mu \nu} (1 + \gamma_5) d_{\alpha}$$
$$J^\mu_{AV} = {\bar u}^{\alpha} \gamma^\mu (1-\gamma_5)
d_{\alpha}$$
 $\alpha$
 is a color index and $\sigma^{\mu
\nu} = (i/2)[\gamma^\mu , \gamma^\nu ]$

\begin{eqnarray}
\nonumber
 \eta_{PS} &=&  \eta_{\chi\tilde e} + \eta_{\chi\tilde f}
+
\eta_{\chi} + \eta_{\tilde g} + 7 \eta_{\tilde g}^{\prime}, \\
\nonumber
  \eta_{T} &=& \eta_{\chi} - \eta_{\chi\tilde f} +
\eta_{\tilde g} - \eta_{\tilde g}^{\prime}
\end{eqnarray}
%
 \begin{eqnarray}
\nonumber \eta_{PS} &=&  \eta_{\chi\tilde e} + \eta_{\chi\tilde f}
+
\eta_{\chi} + \eta_{\tilde g} + 7 \eta_{\tilde g}^{\prime}, \\
\nonumber \eta_{T} &=& \eta_{\chi} - \eta_{\chi\tilde f} +
\eta_{\tilde g} - \eta_{\tilde g}^{\prime}
\end{eqnarray}
with
\begin{eqnarray}
\nonumber \eta_{\tilde g} &=& \frac{\pi \alpha_s}{6}
\frac{\lambda^{'2}_{111}}{G_F^2 m_{\tilde d_R}^4}
\frac{m_P}{m_{\tilde g}}\left[ 1 + \left(\frac{m_{\tilde
d_R}}{m_{\tilde u_L}}\right)^4\right]
 \label{susy1}
 \end{eqnarray}
\begin{eqnarray}\nonumber \eta_{\chi} &=& \frac{ \pi \alpha_2}{2}
\frac{\lambda^{'2}_{111}}{G_F^2 m_{\tilde d_R}^4}
 \nonumber\\
 & & \sum_{i=1}^{4}\frac{m_P}{m_{\chi_i}}
 \left[ \epsilon_{R i}^2(d) + \epsilon_{L i}^2(u)
\left(\frac{m_{\tilde d_R}}{m_{\tilde u_L}}\right)^4\right]
 \label{susy2}
 \end{eqnarray}
\begin{eqnarray}
 \eta_{\chi \tilde e} &=& 2 \pi \alpha_2
\frac{\lambda^{'2}_{111}}{G_F^2 m_{\tilde d_R}^4}
\left(\frac{m_{\tilde d_R}}{m_{\tilde e_L}}\right)^4 \nonumber\\
& & \sum_{i=1}^{4}\epsilon_{L i}^2(e)\frac{m_P}{m_{\chi_i}},
 \label{susy3}
 \end{eqnarray}
 \begin{eqnarray}
  \eta'_{\tilde g} &=& \frac{\pi \alpha_s}{12}
\frac{\lambda^{'2}_{111}}{G_F^2 m_{\tilde d_R}^4}
\frac{m_P}{m_{\tilde g}} \left(\frac{m_{\tilde d_R}}{m_{\tilde
u_L}}\right)^2,
 \label{susy4}
  \end{eqnarray}
 \begin{eqnarray}
  \eta_{\chi \tilde f} &=& \frac{\pi \alpha_2 }{2}
\frac{\lambda^{'2}_{111}}{G_F^2 m_{\tilde d_R}^4}
\left(\frac{m_{\tilde d_R}}{m_{\tilde e_L}}\right)^2
 \nonumber\\
 & &\sum_{i=1}^{4}\frac{m_P}{m_{\chi_i}} [\epsilon_{R i}(d)
\epsilon_{L i}(e)  + \epsilon_{L i}(u) \epsilon_{R i}(d)
\nonumber\\& & (\frac{m_{\tilde e_L}}{m_{\tilde u_L}})^2 +
\epsilon_{L i}(u) \epsilon_{L i}(e) (\frac{m_{\tilde
d_R}}{m_{\tilde u_L}})^2]
 \label{susy5}
\end{eqnarray}
where $$\alpha_2 = g_{2}^{2}/(4\pi)~~,~~\alpha_s =
g_{3}^{2}/(4\pi)$$
 are $SU(2)_L$ and $SU(3)_c$ gauge coupling
constants.
 $$m_{\tilde u_L}~~,~~ m_{\tilde d_R}~~,~~m_{\tilde g}~ and~
m_{\chi}$$
 are masses of the u-squark, d-squark, gluino $\tilde g$
and of the lightest neutralino $\chi$, respectively.
 $$\chi =
Z_{\tilde{B}}\tilde{B} + Z_{\tilde{W}} \tilde{W}^{3} +
Z_{\tilde{H1}}\tilde{H}_{1}^{0} + Z_{\tilde{H2}}
\tilde{H}_{2}^{0}$$
 Here $\tilde{W}^{3}$ and $\tilde{B}$ are
neutral $SU(2)_L$ and $U(1)$ gauginos while $\tilde{H}_{2}^{0}$,
$\tilde{H}_{1}^{0}$ are higgsinos which are the superpartners of
the two neutral Higgs boson fields $H_1^0$ and $H_2^0$.
 The neutralino couplings are defined as
$\epsilon_{L\psi} = - T_3(\psi)Z_{\tilde{W}} + \tan \theta_W
\left(T_3(\psi) - Q(\psi)\right) Z_{\tilde{B}}$, $\epsilon_{R\psi}
= Q(\psi) \tan \theta_W Z_{\tilde{B}}$. Here $Q $ and $ \ T_3$ are
the electric charge and the weak isospin of the fields $\psi = u,
d, e$.

    The lifetime for neutrinoless double beta decay now becomes:
    \begin{eqnarray}
\big[ T_{1/2}^{0\nu} \big]^{-1} &=& G_{01} | \eta_{T} {\cal
M}_{\tilde q}^{2N}
 +  (\eta_{PS}-\eta_{T}) {\cal M}_{\tilde f}^{2N}
 \nonumber\\
 & & + \frac{3}{8}
(\eta_{T} + \frac{5}{8} \eta_{PS}) {\cal M}^{\pi N}  +\eta_N {\cal
M}_N  |^2\,
 \nonumber
\end{eqnarray}
where
\begin{eqnarray}
{\cal M}^{2N}_{\tilde q} &=&  c_A \Big[ \alpha^{(0)}_{V-\tilde{q}}
{\cal M}_{F N} + \alpha^{(0)}_{A-\tilde{q}} {\cal M}_{GT N} +
\nonumber\\
& &\alpha^{(1)}_{V-\tilde{q}} {\cal M}_{F'}
\alpha^{(1)}_{A-\tilde{q}} {\cal M}_{GT'} +
\alpha_{T-\tilde{q}} {\cal M}_{T'} \Big]\,, \nonumber \\
{\cal M}^{2N}_{\tilde f} &=& c_A \Big[ \alpha^{(0)}_{V-\tilde{f}}
{\cal M}_{F N} + \alpha^{(0)}_{A-\tilde{f}} {\cal M}_{GT N} +
\nonumber\\
& & \alpha^{(1)}_{V-\tilde{f}} {\cal M}_{F'}
\alpha^{(1)}_{A-\tilde{f}} {\cal M}_{GT'}
  +\alpha_{T-\tilde{f}} {\cal M}_{T'} \Big]\,, \nonumber \\
{\cal M}^{\pi N} &=& c_A \Big[
 \frac{4}{3}\alpha^{1\pi}\left(M_{GT-1\pi} + M_{T-1\pi} \right)
 \nonumber\\
 & &  +      \alpha^{2\pi}\left(M_{GT-2\pi} + M_{T-2\pi} \right)\Big]\,,
\nonumber \\
{\cal M}_N &=&  c_A \Big[ \frac{g^2_V}{g^2_A} {\cal M}_{F N} -
{\cal M}_{GT N} \Big], c_A = \frac{m_{_p}}{
m_e}\left(\frac{m_A}{m_{_p}}\right)^2.
 \nonumber
\end{eqnarray}
 For comparison we write again the amplitude $\eta_N {\cal
M}_N  |^2$ associated with the heavy neutrino. The other terms
come from SUSY, the first two correspond to the two nucleon mode,
while the third term ${\cal M}^{\pi N}$ corresponds to the pion
mode. The pion mode is dominant \cite{Ver02}, \cite{FKSS97},
recently confirmed with a different method \cite{PREZEAU}. One gets:\\
 $\alpha^{1\pi} = -0.044~~,~~\alpha^{2\pi} = 0.2$ (Elementary particle
 treatment \cite{FKSS97})\\
$\alpha^{1\pi} = -0.012~~,~~\alpha^{2\pi} = 0.15$ (Quark model \cite{Ver02})\\
 Using these ME one obtains from the data the limits:
$$\eta_{N} \leq  \frac{1}{|{\cal M}_N |} \frac{1}{\sqrt{G_{01}
T^{exp}_{1/2}}}$$
 $$\eta_{_{SUSY}} \equiv  \frac{3}{8}(\eta^T + \frac{5}{8}
\eta^{PS})
 \leq  \frac{1}{|{\cal M}^{\pi N}|}
\frac{1}{\sqrt{G_{01} T^{exp}_{1/2}}}$$
\begin{figure}
 \hspace*{-0.0cm}
\includegraphics[height=.3\textheight]{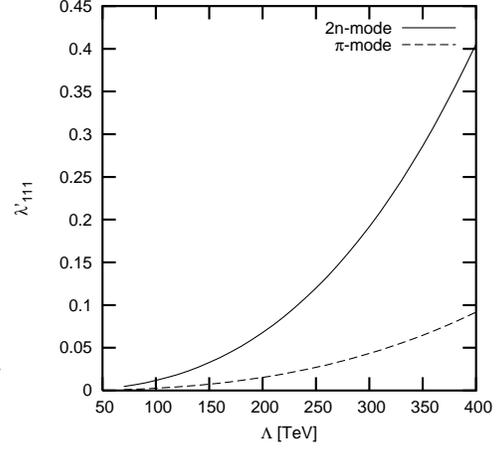}
\caption{The R-parity violating parameter $\lambda^{'}_{111}$ as a
function of the SUSY scale associated with the nucleon and the
pion modes (Taken from Gozdz {\it et al} \cite{KAMINSKI}.
 \label{kaminski1}}
 \end{figure}
 \begin{figure}
 \includegraphics[height=.3\textheight]{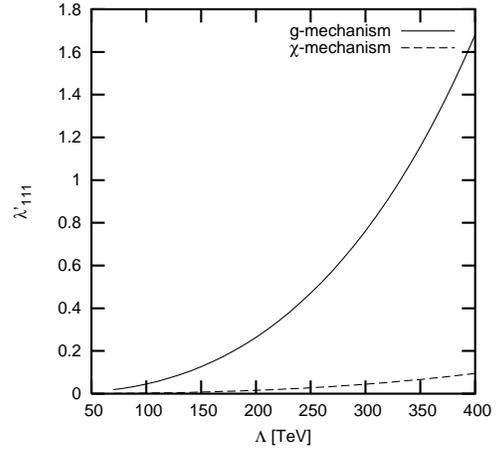}
\caption{The same as in Fig. \ref{kaminski1} in the case of the
pion mode for gluino ($\tilde{g}$) and neutralino
($\tilde{\chi}$).
 \label{kaminski2}}
\end{figure}

 By employing
 standard techniques and working in the allowed SUSY parameter
 space one can obtain  the masses and couplings involving
 the intermediate SUSY particles (gluinos, neutralinos, s-quarks
 and s-leptons) and their couplings, which  enter the expression
  for $\eta_{_{SUSY}}$. Thus from the limits on this
  lepton violating parameter, one can constrain and obtain limits
   on the unknown R-parity violating parameter
 $\lambda^{'}_{111}$.
\section{BRANE WORLD}
It must have become clear from the above discussion that $0\nu
\beta \beta$ decay has something to say in any model, which
predicts massive Majorana neutrinos. We cannot possibly discuss
all such models. In models involving extra dimensions, however,
one has the special feature that they predict Dirac neutrinos.
Furhermore such models predict sterile neutrinos with a mass in
the keV-MeV region. In such cases the particle and nuclear physics
get entangled. So the effective neutrino mass extracted may depend
on the nucleus.
\begin{table}[t]
\begin{center}
\begin{tabular}{|l|r|c|c|c|c|}
\hline
& \multicolumn{5}{c|}{}  \\[-0.2cm]
$1/a$  & \multicolumn{5}{c|}{$\langle m \rangle$~[eV]} \\[0.2cm]
\hline
 & & & & &   \\[-0.2cm]
[GeV] & $^{76}$Ge & $^{100}$Mo & $^{116}$Cd &
$^{130}$Te & $^{136}$Xe   \\[0.2cm]
\hline 0.05  & 0.009 &  0.016 & 0.012 & 0.008 & 0.004  \\
0.1   & 0.052 &  0.061 & 0.062  & 0.050 & 0.025 \\
0.2   & 0.096  &  0.109 & 0.114  & 0.094 & 0.058  \\
0.3   & 0.123   &  0.136 & 0.143 & 0.121 & 0.082 \\
1     & 0.271 &  0.280 & 0.287 & 0.269 & 0.241 \\
10    & 0.493 &  0.494 & 0.495 & 0.493 & 0.489
                                                  \\
\hline
10$^2$ & \multicolumn{5}{c|}{0.513} \\
10$^3$& \multicolumn{5}{c|}{0.535} \\
10$^4$& \multicolumn{5}{c|}{0.066} \\
10$^{10}$ & \multicolumn{5}{c|}{$\stackrel{<}{{}_\sim} 10^{-6}$}\\
\hline
\end{tabular}
\end{center}
\caption{\em Numerical estimates of $\langle m \rangle$ for
different nuclei in a 5-dimensional brane-shifted model.
 (For choice of parameters and further
details see \cite{Pilaftsis} }
 \nonumber
\end{table}
\section{NUCLEAR MATRIX ELEMENTS}
 The discussion of the nuclear matrix elements is outside the
 scope of this talk. So we will limit ourselves to some brief
 comments.
  The first step towards their evaluation consists
  in deriving the effective transition
  operator. Particle physics dictates the structure of the
  operator at the quark level. The next step is going from the quark to the nucleon
  level. This is quite straightforward in the case of light  neutrinos. Since,
  however, the momenta of the intermediate neutrinos are, by
  nuclear standards, quite high, $\approx 70MeV/c$, one must
   consider corrections to the structure of the nucleon current
   \cite{Ver02}
  (induced pseudoscalar etc), which tend  to decrease the nuclear $|ME|^2$
   by $20-30\%$. This causes a corresponding increase in the extracted
  neutrino mass, regardless of the nuclear model. For heavy
  intermediate particles the situation is worse. The effective operator is short ranged.
  So, if one considers only nucleons, the nucleon must not be point like, but as a bag of quarks or
  with a suitable form factor. Other mechanisms must also be
  considered, e.g. $0\nu \beta \beta$ decay induced by pions in flight
  between nucleons (pion mode). This mechanism tends to dominate
  in the case of SUSY contributions.
   The next step is to obtain the needed nuclear wave functions.
   This is a formidable task, since the nuclear systems
   undergoing $0\nu \beta \beta$ decay have complicated structure.
   Furthermore the obtained nuclear matrix elements are quite
   small compared to a canonical value (sum rule) like
   $\sum_f|ME(i \rightarrow f)|^2$
    (the summation is over all
    final states, not just those energetically
    allowed). So normally small effects maybe important here.
     The basic methods employed in the evaluation of these matrix
     elements \cite{Ver02}, \cite{Ver86} are:
     \begin{enumerate}
     \item The Quasiparticle Random Phase Approximation (QRPA).
     One builds the needed intermediate
     states as  RPA excitations on a quasiparticle vacuum
     (one for each ground state). One can include a large
     number of single particle states, since the number of the resulting
     configurations is manageable.
     This method has been applied in almost all systems. Various
     refinements have been incorporated (proton
     neutron-pairing, Pauli principle corrections, $g_{pp}$ renormalization
     factors etc). The thus obtained matrix elements (ME) by various authors
      still have a spread larger than a factor of 2.
      \item The Large Basis Shell Model.
      With modern computers it has been possible to use the
      nuclear shell model in obtaining the nuclear wave function.
      One uses a single particle model space more
      restricted than that employed in the QRPA, since soon the number of
      resulting configurations becomes prohibitively large. With increasing interest
      in $0\nu \beta \beta$ decay we expect the
      nuclear theory groups will improve on this in the
      remainder of the decade, forming consortia, if necessary.
\end{enumerate}
      We thus hope that by the time the ongoing and planned
      experiments give the anticipated results the nuclear matrix
      elements will be known with an accuracy better than a factor
      of 2. The reliability of these matrix elements can be
      tested,
      once the experimental results become available, by taking
      ratios, which, assuming that one mechanism dominates,
       are independent of the lepton violating
      parameters (ratios of lifetimes of different nuclei or
      ratios of transitions to different final states of the same
      nucleus).
 \section{RESULTS}
 $0\nu \beta \beta$ decay is our best hope to set the scale for
neutrino mass. Based on the available experimental results and
using the range of uncertainty in the nuclear matrix elements the
following limits emerge:
 $$ <m_\nu > ~~~<
0.45 - 1.4 eV ~~[^{76}Ge],
              0.7 - 1.5 eV [^{128}Te]$$
$$\lambda  <  (0.8-2.1)\times 10^{-6}~~[^{76}Ge],
                (2.5-7.3)\times 10^{-6}~~[^{136}Xe,]$$
$$\eta <(0.4-1.8)\times 10^{-8}[^{76}Ge ],
              (0.9-3.7)\times 10^{-8}~~[^{128}Te]$$
        Other mechanisms may dominate $0\nu \beta \beta$ decay. One
        popular scenario is to be mediated not by neutrinos, but by SUSY particles.
        From $\eta_{SUSY}$ for each target in the order of table \ref{table.9} we extract the
         value of $\lambda^{'}_{111}$:
         $$(1.0,1.8,1.8,0.8,1.8,0.9,2.3)\times10^{-3}$$
         We should not forget, however, that these values depend
         on the SUSY parameter space, in particular the SUSY
         scale \cite{KAMINSKI}, see Figs \ref{kaminski1} and \ref{kaminski2}.

         Even if the neutrinos do not dominate in $0\nu \beta \beta$
         decay, its observation will demonstrate that they are
         massive Majorana particles.

\begin{table}[htb]
\footnotesize \caption{Summary of lepton violating parameters.}
\bigskip
\label{table.9}
\begin{tabular}{lrrrrr}
\hline \hline
 & $\langle m_{\nu}\rangle$ & $\langle \lambda \rangle$ &
   $\langle \eta \rangle$ & $\langle \eta _N \rangle$ &
   $\langle \eta _{SUSY} \rangle$ \\
\hline
 (A,Z) & eV & $\times10^{-6}$ & $\times10^{-8}$ & $\times10^{-8}$ &
              $\times10^{-8}$  \\
\hline
  & Pr & P1 & P1 & Pr & Pr  \\
\hline
$^{76}Ge$  & 0.51 & 0.97 & 0.55 & 0.76 & 0.54 \\
$^{100}Mo$ & 2.9 & 26 & 8.8 & 6.2 & 3.2  \\
$^{116}Cd$ & 5.9 & 37 & 26 & 11 & 7.6  \\
$^{128}Te$ & 1.8 & 5.6 & 1.3 & 2.9 & 1.6  \\
$^{130}Te$ & 13 & 7.6 & 5.2 & 20 & 10\\
$^{136}Xe$ & 49 & 2.1&1.4 & 4 5 & 2.4  \\
$^{150}Nd$ & 8.5 & 5.6 & 5.3 & 16 & 7.3  \\
\hline \hline
\end{tabular}
\end{table}

\end{document}